\documentclass[12pt]{article}
 \usepackage{pstricks}
  \usepackage[dvips,final]{graphicx}
   \usepackage{amsmath}
    \usepackage{amssymb}
     \usepackage{pifont}   
      \usepackage{psfrag}
       \usepackage{bm}
       
\newrgbcolor{violet}{0.4 0 0.9}                                    
\newcommand{\Ds}{\displaystyle}                                    
\newcommand{\va}[1]{\langle{#1}\rangle}                            
\newcommand{\gev}[1]{\relax\ifmmode{\text{GeV}^{#1}}               
                     \else{{GeV}$^{#1}${ }}\fi}                    
\newcommand{\Gev}{\relax\ifmmode{\text{GeV}}\else{{ GeV}{ }}\fi}   

\begin{document}

 \begin{center}
 \textbf{\LARGE  Pion distribution amplitude and form-factors:\\[1mm]
  Improved Gaussian model of QCD vacuum}\\[5mm]
 
 A.~P.~Bakulev\footnote{E-mail: bakulev@theor.jinr.ru}\\[3mm]

\textit{Joint Institute for Nuclear Research,\\
        Bogoliubov  Lab. of Theoretical Physics,\\
        141980, Moscow Region, Dubna, Russia}\\[0.9cm]
\end{center}\vspace*{-5mm}

\begin{abstract}
 We describe the present status of the pion distribution amplitude (DA)
 as it originated from several sources:
 (i) a nonperturbative approach, based on QCD sum rules with nonlocal
 condensates,
 (ii) an $O(\alpha_s)$ QCD analysis of the CLEO data on
 $F^{\gamma\gamma^*\pi}(Q^2)$
 with asymptotic and renormalon models for higher twists, and
 (iii) recent high-precision lattice QCD calculations 
 of the second moment of the pion DA.
 Then we show the comparison of the results 
 for the pion electromagnetic form factor,
 obtained in analytic perturbation theory,
 with JLab data on $F_{\pi}(Q^2)$.
 After that we introduce the improved model for nonlocal condensates 
 in QCD and show its consequences for the pion DA and $\gamma\gamma^*\to\pi$
 transition form factor.
 In order to facilitate possible applications of BMS and improved ``bunches''
 we suggest approximate analytic descriptions of their boundaries.
\end{abstract}

\section{Generalized QCD SRs for mesonic distribution amplitudes}
The pion distribution amplitude (DA) parameterizes
the matrix element of the nonlocal axial current on the light cone~\cite{Rad77}
\begin{eqnarray}
 \label{eq:pion.DA.ME}
 \va{0\!\mid\!\bar d(z)\gamma_{\mu}\gamma_5 
  E(z,0)u(0)\!\mid\!\pi(P)}\Big|_{z^2=0}
  = i f_{\pi}P_{\mu}\!
       \int\limits_{0}^{1}\! dx\ e^{ix(zP)}
        \varphi_{\pi}^\text{Tw-2}(x,\mu^2)\,.~~
\end{eqnarray}
The gauge-invariance of this DA is guarantied by the Fock--Schwinger string
$$E(z,0)={\cal P}\exp\left[i g\int_0^z A_\mu(\tau) d\tau^\mu\right]$$,
inserted in between separated quark fields.
Physical meaning of this DA is quite evident: it is the amplitude 
for the transition $\pi(P)\rightarrow u(Px) + \bar{d}(P(1-x))$.
It is convenient to represent the pion DA using expansion 
in Gegenbauer polynomials $C^{3/2}_n(2x-1)$,
which are one-loop eigenfunctions of ERBL kernel~\cite{ER80,LB79}
\begin{eqnarray}
 \label{eq:pi.DA.rep}
  \varphi_\pi(x;\mu^2) 
  = \varphi^\text{As}(x)\,
     \Bigl[1 + \sum\limits_{n\geq1}a_{2n}(\mu^2)\,C^{3/2}_{2n}(2x-1)
     \Bigr]\,,
\end{eqnarray}
where $\varphi^\text{As}(x)=6\,x\,(1-x)$ is the famous asymptotic DA.
This representations means 
that all scale dependence  in $\varphi_\pi(x;\mu^2)$
is transformed to the scale dependence of the set 
$\left\{a_2(\mu^2), a_4(\mu^2), \ldots\right\}$.
We mention here 
that ERBL solution at the 2-loop level
is also possible 
with 
using the same representation 
(\ref{eq:pi.DA.rep})~\cite{MR86ev,KMR86,Mul94,BS05}.

In order to construct reliable QCD SRs for the pion DA
moments 
one needs, as has been shown in~\cite{MR86,BM98},
to take into account 
the nonlocality of QCD vacuum condensates.
For an illustration of the nonlocal condensate (NLC) model
we use here the minimal Gaussian model
\begin{eqnarray}
 \label{eq:Min.Gauss.Mod}
  \va{\bar{q}(0)q(z)} 
   = \va{\bar{q}\,q}\,
      e^{-|z^2|\lambda_q^2/8}
\end{eqnarray} 
with a single scale parameter $\lambda_q^2 = \va{k^2}$,
which characterizes 
the average momentum of quarks in the QCD vacuum.
Its value has been estimated in QCD SR approach 
and also on the lattice\cite{BI82lam,OPiv88,DDM99,BM02}:
\begin{eqnarray}
 \label{eq:lambda.q.SR}
  \lambda_q^2
   = 0.35-0.55~\gev{2}\,.
\end{eqnarray}  
We see that $\lambda_q^2$ is of an order of the typical hadronic scale
~$m_{\rho}^2 \approx 0.6$ GeV$^2$. 

Let us write down as an example 
the NLC QCD SR for the pion DA $\varphi_{\pi}(x)$.
To produce it one starts 
with a correlator of currents
$J_{\mu5}(x)$ and 
$J^{\dagger}_{\nu5;N}(0)=\bar{d}(0)\,\hat{n}\,\gamma_5\left(n\nabla\right)^N\!u(0)$
with light-like vector $n$, $n^2=0$,
obtains SRs for the moments $\va{x^N}_{\pi}$
and then applies the inverse Mellin transform,
$\va{x^N}_{\pi} \Rightarrow \varphi_\pi(x)$.
As a result we obtain
\begin{eqnarray}
 \label{eq:NLC.SR.pion.DA}
   f_{\pi}^2\,\varphi_\pi(x) 
   = \int_{0}^{s_{0}}\!\!\rho^\text{pert}(x;s)\,
         e^{-s/M^2}ds 
     + \frac{\alpha_s\langle GG\rangle}{24\pi M^2}\,
        \varphi_{GG}(x;\Delta)
     + \frac{8\pi\alpha_s\va{\bar{q}q}^2}{81M^4}
        \sum_{i=2V,3L,4Q}\varphi_i(x;\Delta)
\end{eqnarray}
with $\Delta\equiv\lambda_q^2/M^2$. 
The local limit $\Delta\to0$ of this SR 
is specified by the appearance of $\delta$-functions 
concentrated at the end-points $x=0$ and $x=1$,
for example,
$\varphi_{4Q}(x;\Delta)=9[\delta(x)+\delta(1-x)]$.
\begin{figure}[hb]
 \centerline{%
  \begin{minipage}{0.8\textwidth}
   \centerline{\includegraphics[width=0.625\textwidth]{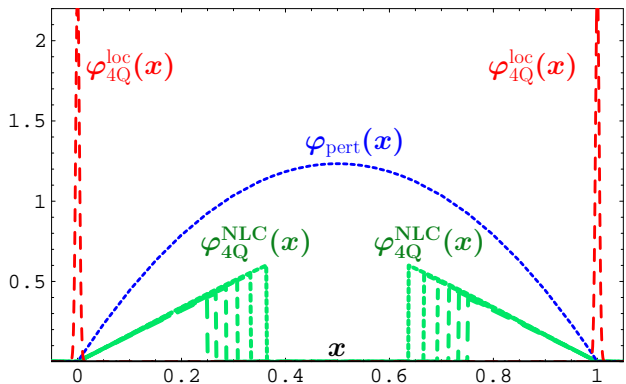}}
   \caption{We show here contributions 
   to the Eq.\ (\ref{eq:NLC.SR.pion.DA}) due to perturbative loop 
   (dotted line) and due to 4Q-condensate: $\varphi^\text{loc}_\text{4Q}(x)$
   --- in standard QCD SRs, and $\varphi^\text{NLC}_\text{4Q}(x,M^2=0.55-0.80~\text{GeV}^2)$
   --- in NLC QCD SRs.
   \label{fig:piDA.SR.4Q}}
 \end{minipage}}
\end{figure}

\noindent
The minimal Gaussian model (\ref{eq:Min.Gauss.Mod})
generates the contribution $\varphi_\text{4Q}(x;\Delta)$
shown in Fig.\ \ref{fig:piDA.SR.4Q}
in comparison with the perturbative one
for the standard (local) and the NLC types of the SR.
We see that due to completely different behaviour 
of perturbative and condensate terms 
in the local QCD SR case 
it is difficult to obtain some kind of consistency.
Alternatively, the NLC contribution is much more similar 
to the perturbative one.
Just by this reason we have very good stability in the NLC SR case!
After processing SR (\ref{eq:NLC.SR.pion.DA}) 
for the moments $\va{\xi^N}_\pi = \int_0^1\varphi_\pi(x)\,\left(2x-1\right)^N dx$,
we restore the pion DA $\varphi_\pi(x)$ 
by demanding that it should reproduce these first five moments
$\va{\xi^i}_\pi$, $i=2\,, 4\,, \ldots\/, 10$
with using the minimally possible number of the Gegenbauer harmonics
in representation (\ref{eq:pi.DA.rep}).
It appears that NLC SRs for the pion DA
produce a bunch of self-consistent 2-parameter models
at $\mu_0^2\simeq 1.35$ GeV$^2$:
\begin{figure}[t]
 \centerline{\begin{minipage}{0.9\textwidth}\hspace*{-0.06\textwidth}
  \begin{tabular}{cc}
   \begin{minipage}{0.52\textwidth}\vspace{-1mm}
    \centerline{\includegraphics[width=\textwidth]{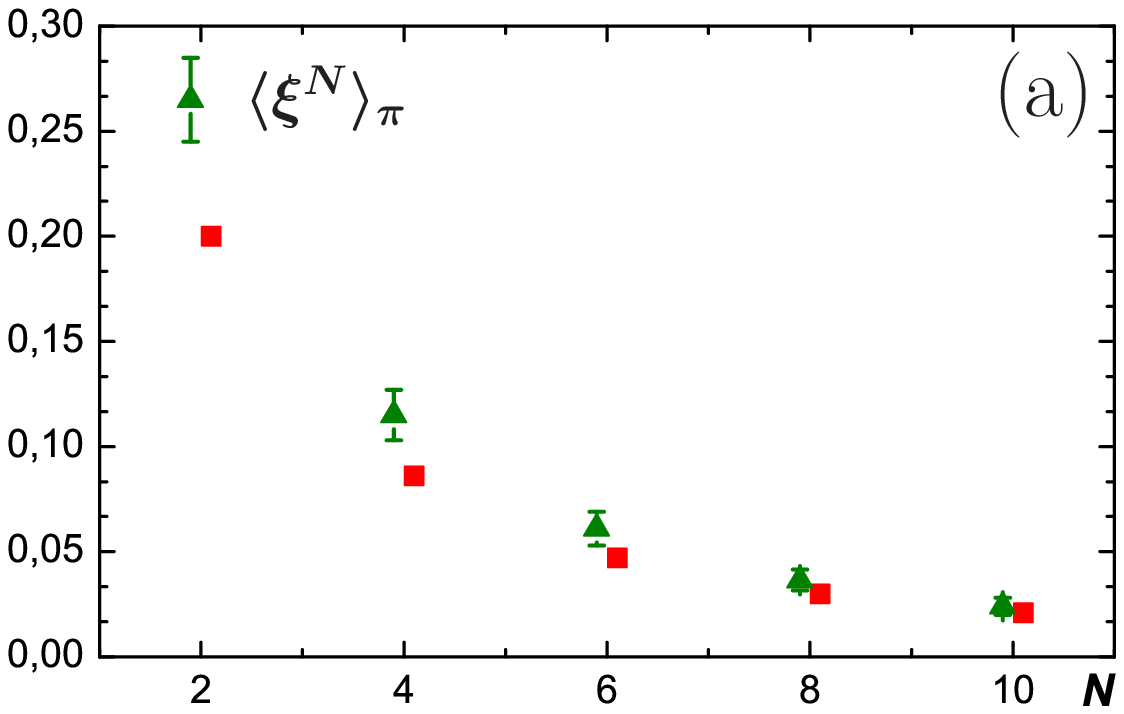}}
   \end{minipage}&
   \begin{minipage}{0.539\textwidth}
    \centerline{\includegraphics[width=\textwidth]{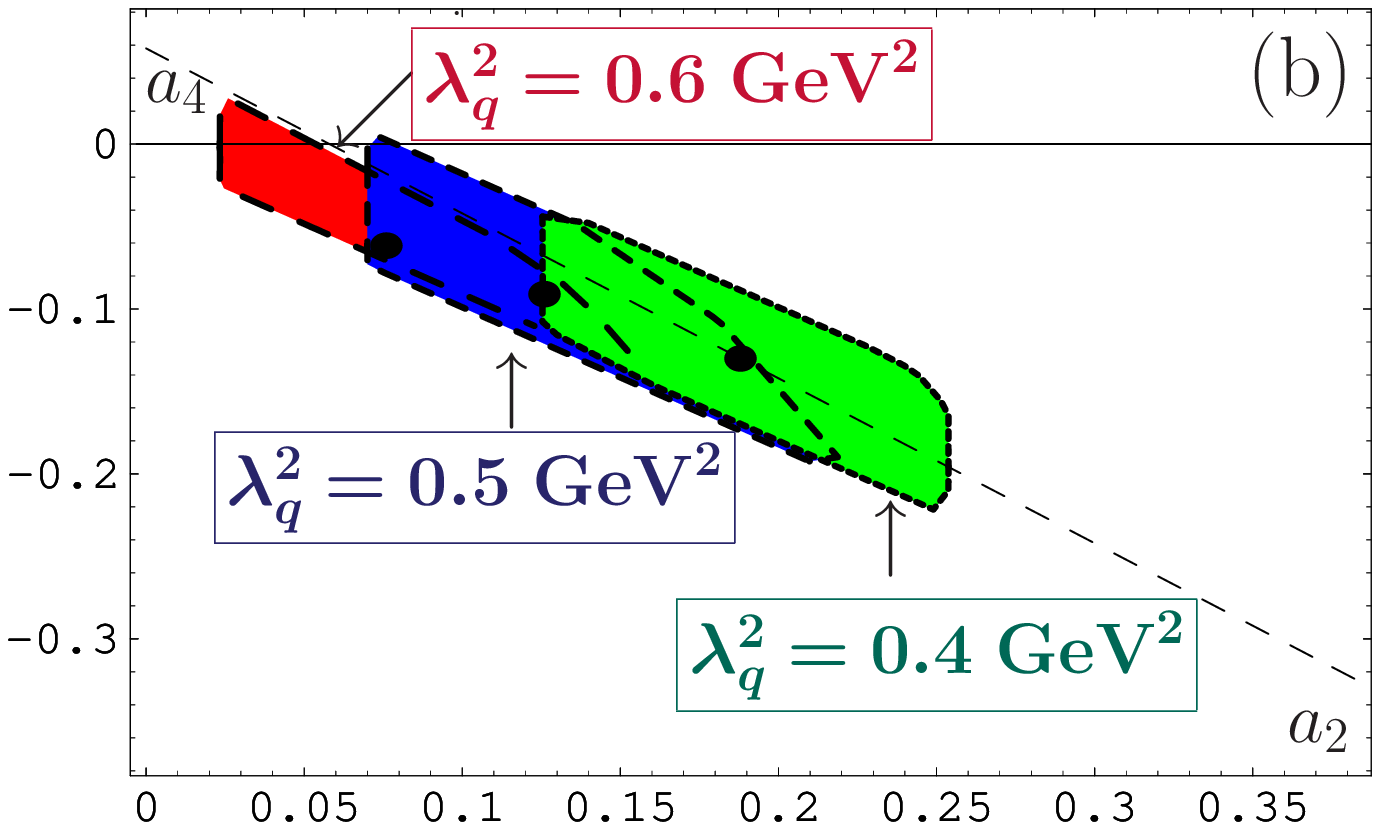}}
   \end{minipage}
  \end{tabular}
   \caption{\textbf{Panel (a)}: 
  Moments $\va{\xi^{N}}_\pi$ with $N=2, \ldots, 10$, 
  obtained using the NLC SR (\ref{eq:NLC.SR.pion.DA}), 
  are shown by triangles with error-bars in comparison with 
  the asymptotic DA moments (squares).
  \textbf{Panel (b)}: The allowed values of parameters $a_2$ and $a_4$ 
  of the bunches (\ref{eq:pi.DA.2Geg}),
  evaluated at $\mu^2=1.35$~GeV$^2$ for three values 
  of the nonlocality parameter $\lambda_q^2=0.4\,, 0.5$, and $0.6$~GeV$^2$.
  \label{fig:456}}
 \end{minipage}}
\end{figure}
\begin{eqnarray}
 \label{eq:pi.DA.2Geg}
  \varphi^\text{NLC}_\pi(x;\mu_0^2) 
  = \varphi^\text{As}(x)\,
     \Bigl[1 + a_{2}(\mu_0^2)\,C^{3/2}_{2}(2x-1)
             + a_{4}(\mu_0^2)\,C^{3/2}_{4}(2x-1)
     \Bigr]\,.~
\end{eqnarray}
The central point corresponds to $a_2^\text{BMS}=+ 0.188$, $a_4^\text{BMS}=-0.130$
in the case where $\lambda^2_q=0.4$ GeV$^2$,
whereas other allowed values of parameters $a_2$ and $a_4$
are shown in the left panel of Fig.\ \ref{fig:456} 
as the slanted rectangle~\cite{BMS01}.
By self-consistency of these solutions
we understand 
that all of them produce,
in accord with (\ref{eq:pi.DA.2Geg}),
the inverse moment of the pion DA
\begin{eqnarray}
 \label{eq:pion.DA.Inv.Mom}
  \va{x^{-1}}^\text{bunch}_{\pi}
   = 3.17\pm0.20\,.
\end{eqnarray}
\begin{figure}[b]\hspace*{-3mm}
 \centerline{%
  \begin{minipage}{0.9\textwidth}
   \centerline{\includegraphics[width=0.625\textwidth]{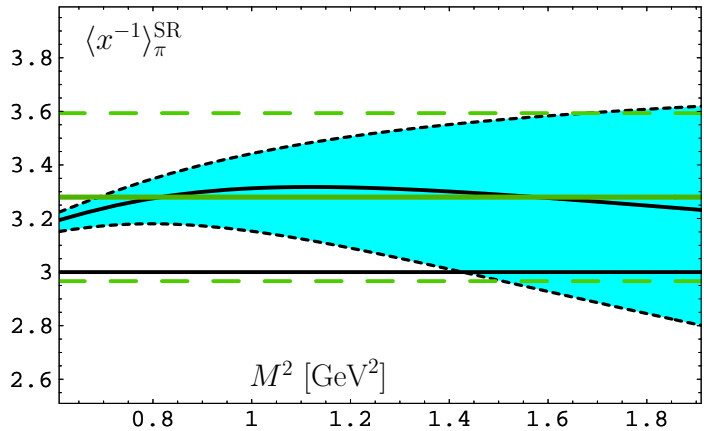}}
    \caption{The inverse moment $\va{x^{-1}}_\pi$, 
    obtained using the NLC SR (\ref{eq:NLC.SR.pion.DA}), 
    is shown by the solid line (central value) with error-bars,
    shown as dashed lines.
    \label{fig:invers}}
 \end{minipage}}
\end{figure}
And this range is in a good agreement
with the estimation
dictated by the special SR for this moment,
which can be obtained through the basic SR (\ref{eq:NLC.SR.pion.DA})
by integration in $x$ 
with the corresponding weight $x^{-1}$ (at $\mu_0^2\simeq 1.35$ GeV$^2$):
\begin{eqnarray}
 \label{eq:Inv.Mom.SR}
 \va{x^{-1}}_{\pi}^{\text{SR}}=3.30\pm0.30\,,
\end{eqnarray}
see Fig.\ \ref{fig:invers}.
It is worth to emphasize here that the moment 
$\va{x^{-1}}^\text{SR}_{\pi}$ could be determined 
only in NLC SRs 
because there are no end-point singularities.

\begin{figure}[h]
 \centerline{\begin{minipage}{0.9\textwidth}\hspace*{-0.08\textwidth}
  \begin{tabular}{cc}
  \begin{minipage}{0.544\textwidth}
   \centerline{\includegraphics[width=\textwidth]{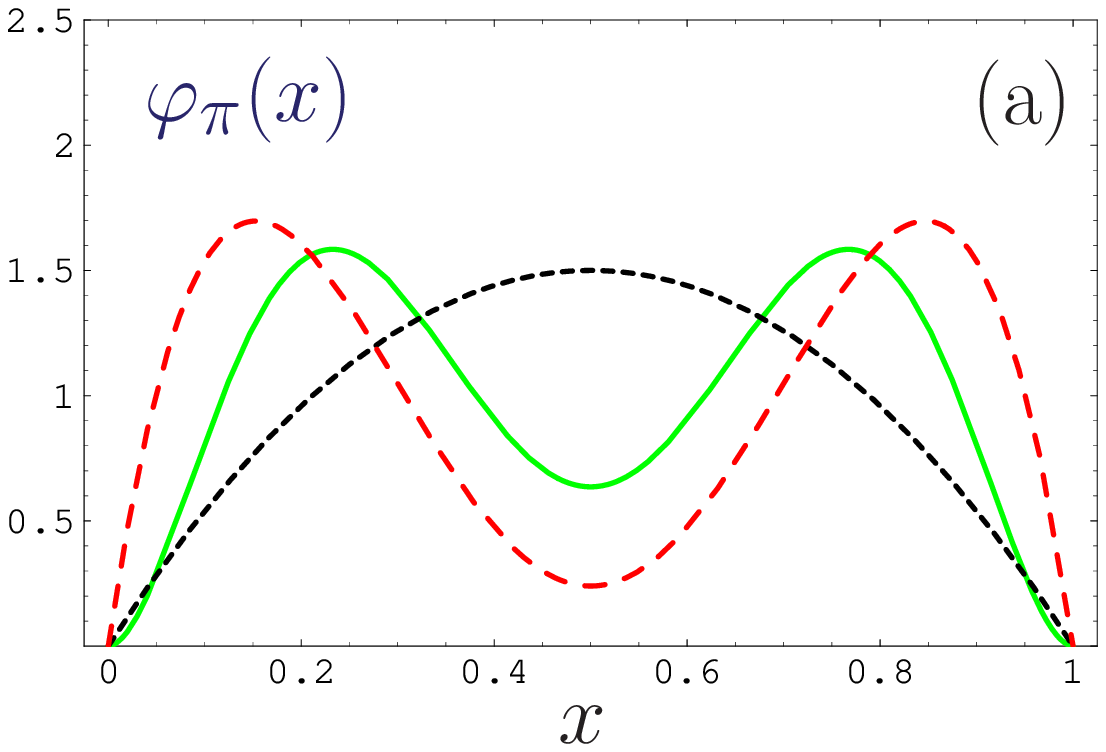}}
  \end{minipage}&
  \begin{minipage}{0.544\textwidth}
   \centerline{\includegraphics[width=\textwidth]{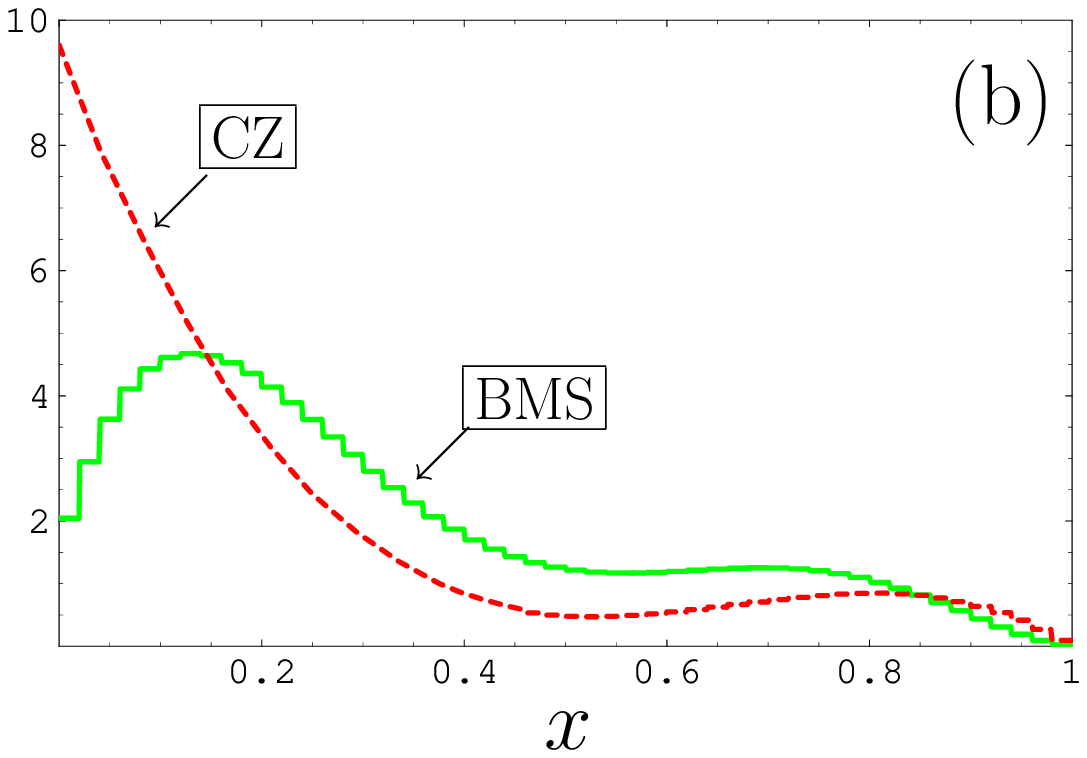}}
   \end{minipage} 
 \end{tabular}
 \caption{\textbf{Panel (a)}: We compare here shapes of three DAs --- 
  BMS (solid line), CZ (dashed line), and the asymptotic DA (dotted line).
  \textbf{Panel (b)}: Histograms for contributions of different bins 
  to inverse moment $\va{x^{-1}}_{\pi}$ are shown for CZ and BMS DAs.
  \label{fig:CZ.As.BMS}}
 \end{minipage}}  
\end{figure}

Comparing the obtained pion DA with 
the Chernyak\&Zhitnitsky (CZ) one~\cite{CZ82}
reveals that although both DAs are two-humped
they are quite different ---
BMS DA is strongly end-point suppressed,
as illustrated in Fig.\ \ref{fig:CZ.As.BMS}(a).
To display this property more explicitly
we show in the panel (b) of this figure
the comparison of BMS and CZ contributions of different bins 
to inverse moment $\va{x^{-1}}_{\pi}$, 
calculated as $\int_x^{x+0.02}\varphi(x) dx$ and normalized to 100\%.

\section{Analysis of CLEO data on $F_{\gamma\gamma^*\pi}(Q^2)$ and pion DA}
Many studies \cite{SY99,SSK99,AriBro-02,BM02,BMS02,Ag05a}
have been performed in the literature 
to determine the pion DA 
using the high-precision CLEO data \cite{CLEO98}
on the pion-photon transition form factor 
$F_{\pi\gamma^{*}\gamma}(Q^2)$. 
In particular, in \cite{BMS02} we have used 
Light-Cone Sum Rules (LCSR) \cite{Kho99,SY99} 
to the next-to-leading-order accuracy 
of QCD perturbation theory 
to examine the theoretical uncertainties 
involved in the CLEO-data analysis 
in order to extract 
more reliably the first two non-trivial Gegenbauer coefficients 
$a_2$ and $a_4$, which parameterize the deviation
from the asymptotic expression $\varphi_{\pi}^\text{As}$.

Why does one need to use Light-Cone SRs (LCSRs) 
in analyzing the experimental data on $\gamma^*(Q)\gamma(q)\to\pi^0$-transition 
form factor?
For $Q^2\gg m_\rho^2$, $q^2\ll m_\rho^2$ 
the QCD factorization is valid only in the leading twist approximation 
and the higher twists are of importance~\cite{RR96}.
The reason is quite clear: 
if $q^2\to0$ one needs to take into account interaction 
of a real photon at long distances 
of order of $O(1/\sqrt{q^2})$.
Then in order to account for long-distance effects
in perturbative QCD 
one needs to introduce the light-cone DA of the real photon.
Instead of doing so,
Khodjamirian~\cite{Kho99} suggested to use the LCSR approach,
which effectively accounts for long-distances effects 
of the real photon using quark-hadron duality 
in the vector channel and dispersion relation in $q^2$.
Schmedding and Yakovlev realized this approach
to analyze the CLEO data on 
the $\gamma^*\gamma\to\pi$ transition form factor
with the next-to-leading-order (NLO) accuracy 
of perturbative QCD part of LCSR~\cite{SY99}.

We improved the NLO analysis of the CLEO data
by taking into account 
the following points: 
(i) the NLO evolution for both {$\varphi(x, Q^2_\text{exp})$} 
    and $\alpha_s(Q^2_\text{exp})$ with accurate taking into account 
    heavy quark thresholds;
(ii) the relation  between the ``nonlocality" scale and 
     the twist-4 magnitude 
     $\delta^2_\text{Tw-4} \approx \lambda_q^2/2$ 
     was used to re-estimate 
     $\delta^2_\text{Tw-4}=0.19 \pm 0.02$ 
     at $\lambda_q^2=0.4$ GeV$^2$;
(iii) the possibility to extract constraints on $\va{x^{-1}}_\pi$ 
      from the CLEO data and 
      to compare them with those we have from NLC QCD SRs.
      
\begin{figure}[ht]
 \centerline{\begin{minipage}{0.93\textwidth}\hspace*{-0.013\textwidth}
 \centerline{\includegraphics[width=0.353\textwidth]{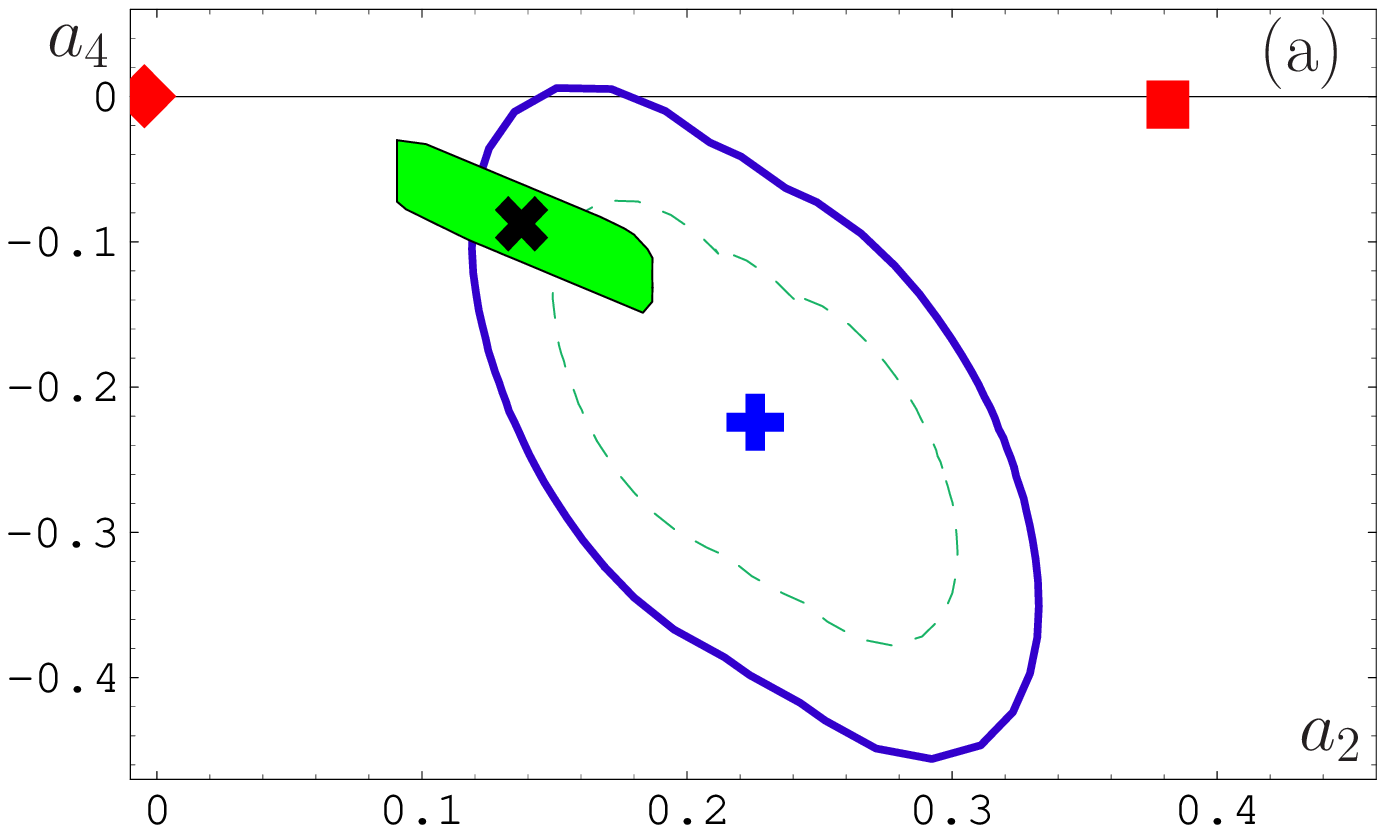}%
            ~\includegraphics[width=0.353\textwidth]{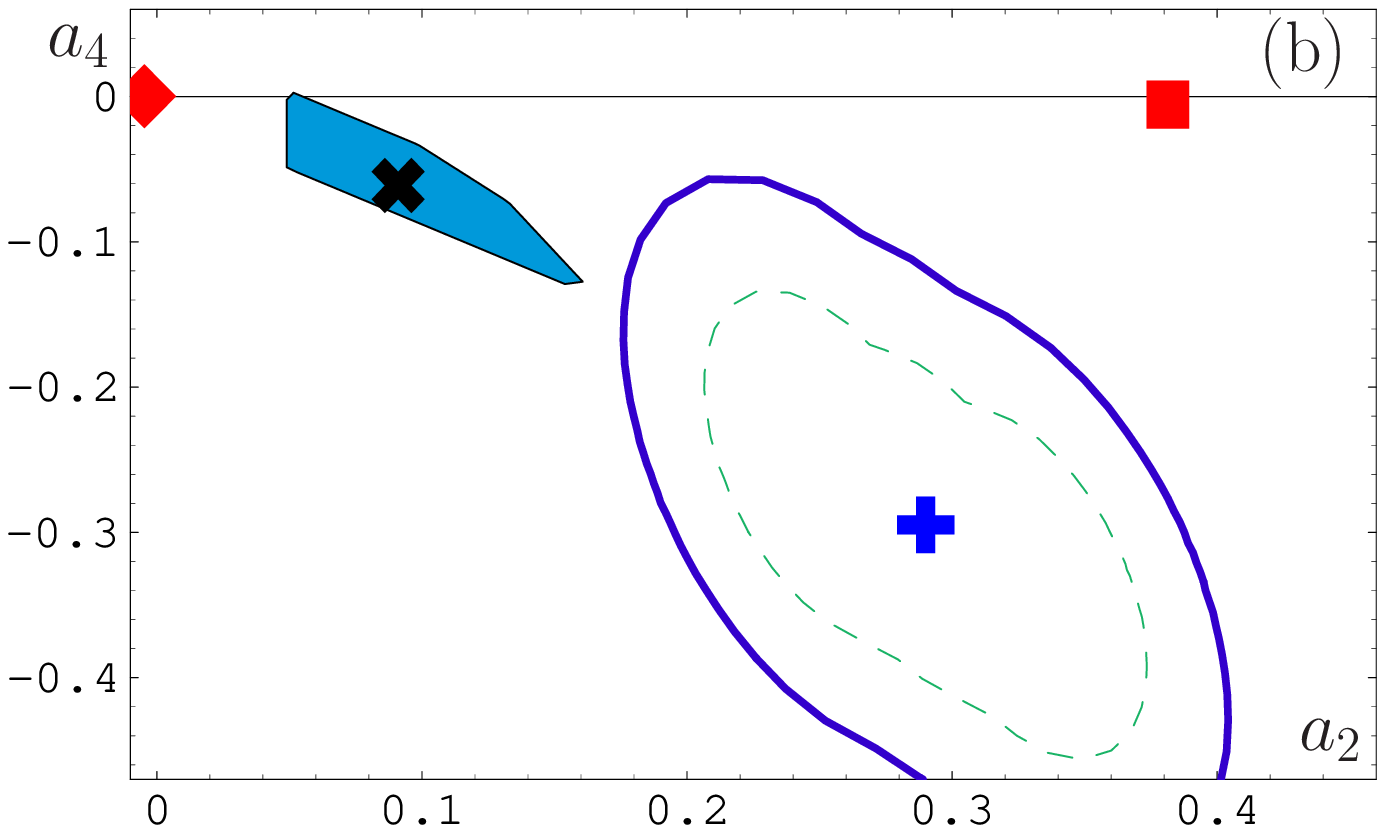}%
            ~\includegraphics[width=0.353\textwidth]{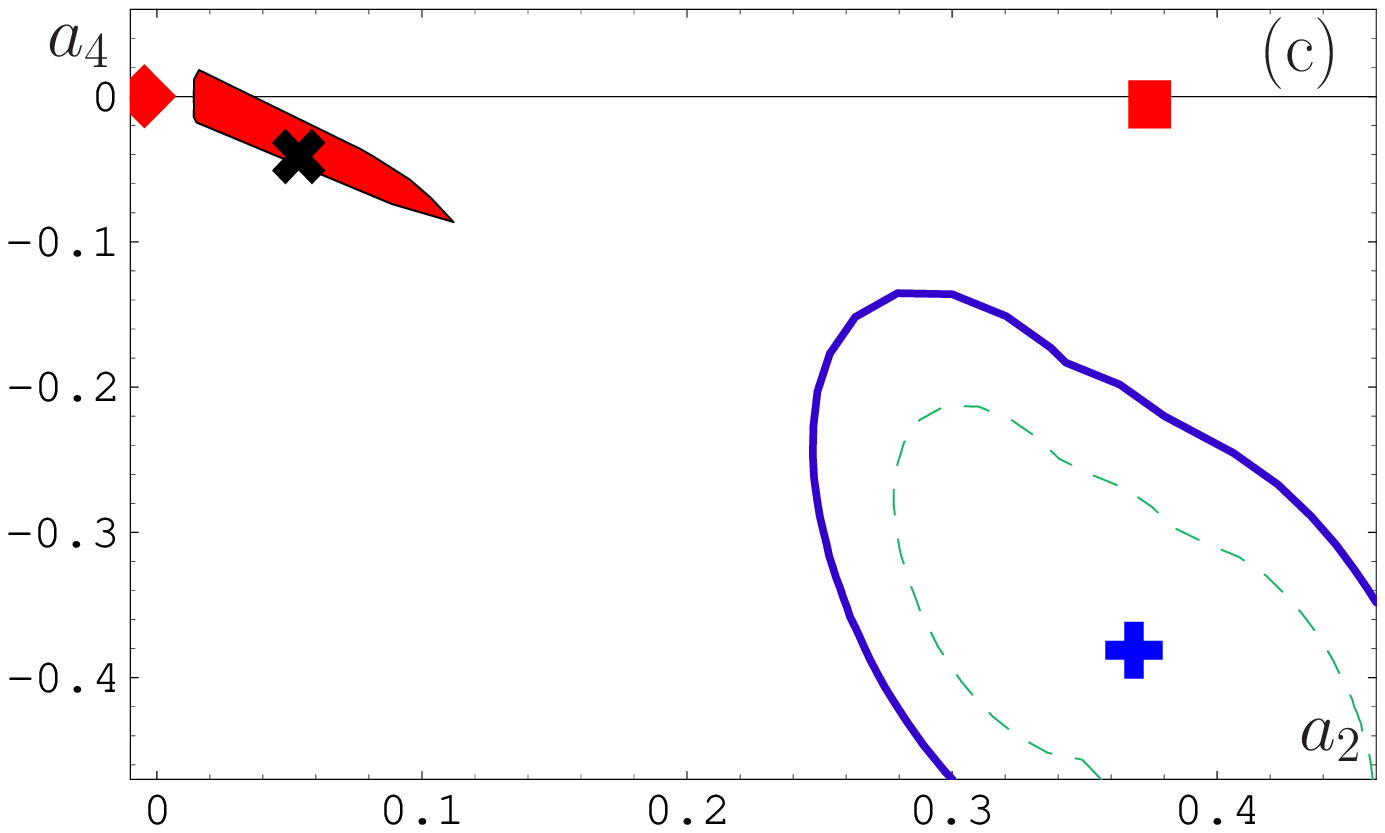}}%
  \caption{Three $2\sigma$- and $1\sigma$-contours (solid and dashed lines, correspondingly) 
   of the admissible regions following from the analysis of the CLEO data 
   for different values of $\delta^2$:
   (a) -- for $\lambda^2_q=0.4~\gev{2}$ and $\delta^2=(0.19\pm0.02)~\gev{2}$;
   (b) -- for $\lambda^2_q=0.5~\gev{2}$ and $\delta^2=(0.235\pm0.025)~\gev{2}$;
   (c) -- for $\lambda^2_q=0.6~\gev{2}$ and $\delta^2=(0.29\pm0.03)~\gev{2}$.
   \label{fig:cleo.456}}
 \end{minipage}}
\end{figure}

Results of our analysis~\cite{BMS02} 
are displayed in Fig.\ \ref{fig:cleo.456}.
Solid lines in all figures enclose the $2\sigma$-contours,
whereas the $1\sigma$-contours are enclosed by dashed lines.
The three slanted and shaded rectangles represent the constraints
on ($a_2,~a_4$) posed by the QCD SRs~\cite{BMS01}
for corresponding values of $\lambda^2_q=0.6,~0.5,~0.4$~\gev{2}
(from left to right).
All values are evaluated at $\mu^2=2.4$~GeV$^2$ after the NLO evolution.

\begin{figure}[h]
 \centerline{\begin{minipage}{0.93\textwidth}\hspace*{-0.013\textwidth}
  \begin{tabular}{cc}
   \begin{minipage}{0.48\textwidth}
    \centerline{\includegraphics[width=\textwidth]{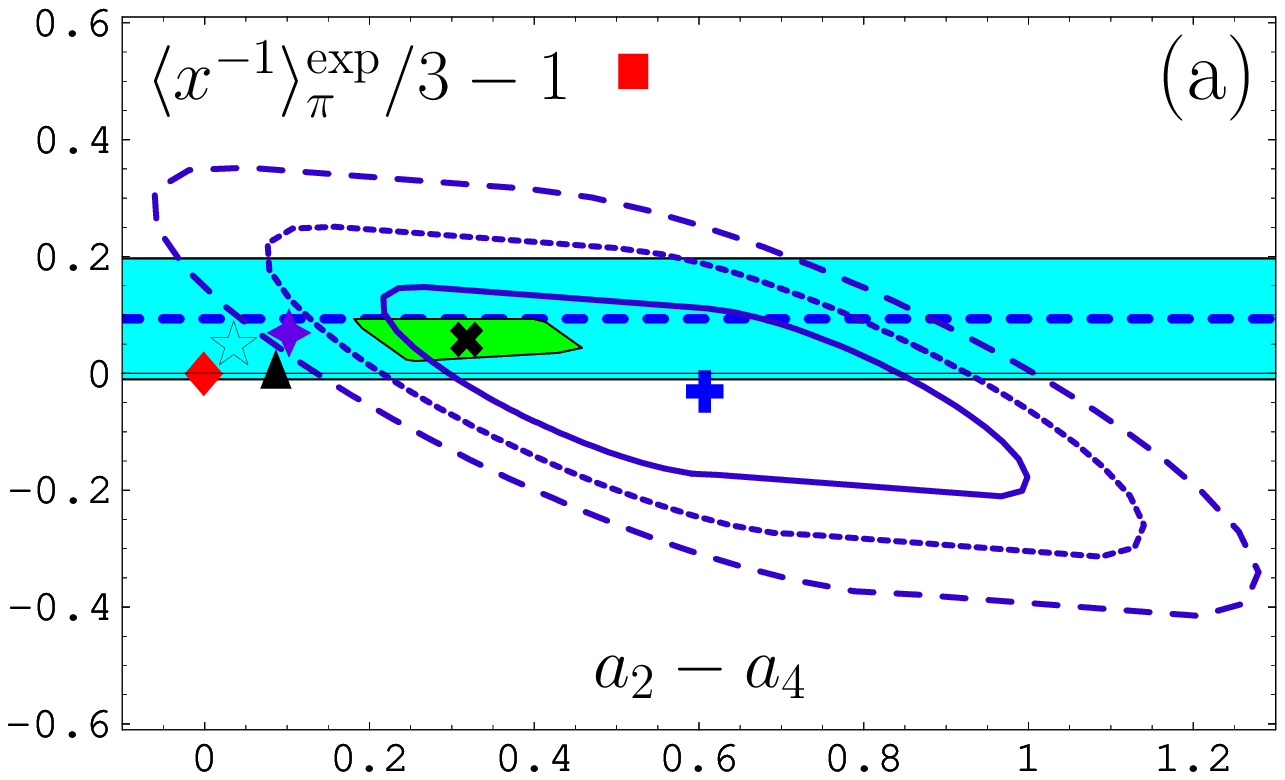}}
   \end{minipage}&
   \begin{minipage}{0.48\textwidth}
    \centerline{\includegraphics[width=\textwidth]{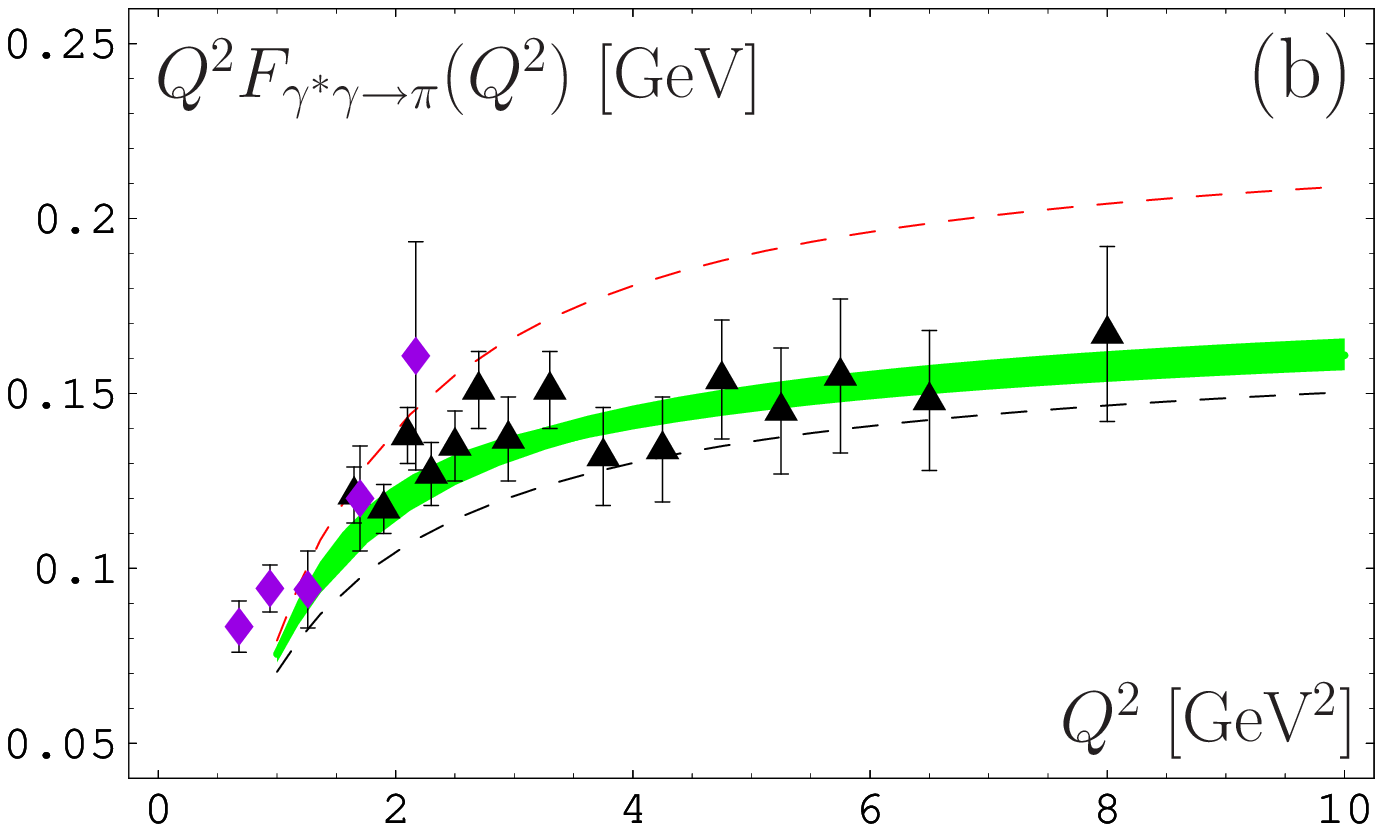}}
   \end{minipage} 
 \end{tabular}
  \caption{\textbf{(a)}: The results of the CLEO data analysis 
    for the pion DA parameters
        ($\Ds \langle x^{-1} \rangle^\text{exp}_{\pi}/3-1$,
    evaluated at $\mu^2_0 \approx 1~\gev{2}$.
    \textbf{(b)}: LCSR predictions for 
    $Q^2F_{\gamma^*\gamma\to\pi}(Q^2)$
    for the CZ DA (upper dashed line),
    BMS-``bunch'' (shaded strip), 
    and the asymptotic DA (lower dashed line)
    in comparison with the CELLO 
   (diamonds, \protect\cite{CELLO91}) and the CLEO 
   (triangles, \protect\cite{CLEO98}) experimental data,
   evaluated with the twist-4 parameter value 
   $\delta_{\rm Tw-4}^2=0.19$~GeV$^2$~\protect\cite{BMS02}
   and at $\mu^2_\text{SY}=5.76~\text{GeV}^2$.
        \label{fig:inv.mom.ff.cello}\vspace*{-1mm}}
 \end{minipage}} 
\end{figure}

We see that the CLEO data definitely prefer the value of
the QCD nonlocality parameter $\lambda_q^2 = 0.4$ GeV$^2$.
We also see in Fig.\ \ref{fig:cleo.456}(c)
(and this conclusion was confirmed even 
 with 20\% uncertainty in twist-4 magnitude, 
 see also Fig.\ \ref{fig:inv.mom.ff.cello}(a))
that CZ DA ({\red\footnotesize\ding{110}}) is excluded at least at $4\sigma$-level, 
whereas the asymptotic DA ({\red\ding{117}}) --- at $3\sigma$-level.
In the same time our DA (\ding{54}) 
and most of the bunch (the slanted green-shaded rectangle around the symbol \ding{54}) 
are inside 1$\sigma$-domain.
Instanton-based Bochum (\ding{73}) and Dubna ($\blacktriangle$) models 
are near 3$\sigma$-boundary
and only the Krakow model~\cite{PR01},
denoted in Fig.\ \ref{fig:inv.mom.ff.cello}(a) 
by symbol {\violet\ding{70}},
is close to 2$\sigma$-boundary.

In Fig.\ \ref{fig:inv.mom.ff.cello}(a) 
we demonstrate the $1\sigma$-, $2\sigma$- and $3\sigma$-contours
(solid, dotted and dashed contours around
 the best-fit point ({\blue\ding{58}})), 
which have been obtained for values of the twist-4 scale parameter 
$\delta_\text{Tw-4}^2=[0.15-0.23]~\gev{2}$.
As one sees from the blue dashed line within the hatched band,
corresponding in this figure to the mean value of 
$\Ds \langle x^{-1} \rangle^\text{SR}_{\pi}/3-1$
and its error bars,
the nonlocal QCD sum-rules result with its error bars
appears to be in good agreement with the CLEO-constraints 
on $\Ds \langle x^{-1} \rangle^\text{exp}_{\pi}$ 
at the $1\sigma$-level,
Moreover, the estimate $\Ds \langle x^{-1} \rangle^\text{SR}_{\pi}$
is close to $\Ds \langle x^{-1} \rangle^\text{EM}_{\pi}/3-1=0.24\pm 0.16$,
obtained in the data analysis of the electromagnetic pion form factor
within the framework of a different LCSR method in \cite{BKM00,BK02}.
These three independent estimates are in good agreement to each other,
giving firm support that the CLEO data processing, on one hand, 
and the theoretical calculations, on the other, 
are mutually consistent.

Another possibility, suggested in~\cite{Ag05b}, 
to obtain constraints on the pion DA in the LCSR analysis 
of the CLEO data  -- 
to use for the twist-4 contribution renormalon-based model,
relating it then to parameters $a_2$ and $a_4$ of the pion DA.
Using this method we obtain~\cite{BMS05lat} the renormalon-based constraints 
for the parameters $a_2$ and $a_4$,
shown Fig.\ \ref{fig:Lat.Ren.CLEO} 
in the form of $1\sigma$-ellipse (dashed contour).

\section{Dijet E791 data, pion form factor and CEBAF data}
\begin{figure}[h]
 \centerline{\begin{minipage}{0.93\textwidth}\hspace*{-0.013\textwidth}
  \centerline{\includegraphics[width=\textwidth]{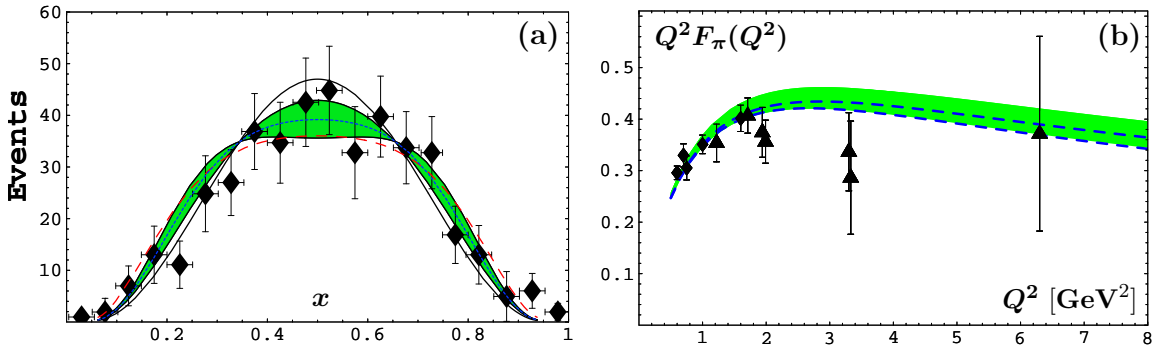}}
   \caption{\textbf{(a)}: Comparison with the E791 data
        on diffractive dijet production of the BMS ``bunch''
        (shaded strip), the asymptotic DA (solid line), and the CZ
        (dashed line) model, using the convolution approach of
        \cite{BISS02}.
    \textbf{(b)}: The scaled pion form factor 
    calculated with the BMS\ bunch (shaded strip) and asymptotic DA (dashed lines)
    including nonperturbative uncertainties from NLC QCD SRs~\protect{\cite{BMS01}}
    and renormalization scheme and scale ambiguities at the $O(\alpha_s^2)$-level~\cite{BPSS04}.
    The experimental data are taken from \protect{\cite{JLAB00}}
    (diamonds) and \cite{FFPI73}, \cite{FFPI76} (triangles).
    \label{fig:dijet.FF.Pi}}
 \end{minipage}}
\end{figure}

Our findings are further confirmed by the E791 data~\cite{E79102} 
on diffractive dijet $\pi+A$-production. 
This is illustrated in Fig.\ \ref{fig:dijet.FF.Pi}(a).
The main conclusion here is that all considered pion DAs
are consistent with the data, 
with tiny preference to the BMS DA.
Indeed, following  the convolution procedure of~\cite{BISS02}
we found~\cite{BMS02} the following values of $\chi^2$ 
for three types of pion DAs:
12.56 (asymptotic DA), 10.96 (BMS bunch), and 14.15 (CZ DA).

It is worth to mention here also the results of our analysis of
the pion electromagnetic form factor
using NLC dictated pion DA and Analytic Perturbative QCD~\cite{BPSS04}.
These results are in excellent agreement
with CEBAF data on pion form factor, 
as illustrated in the Fig.\ \ref{fig:dijet.FF.Pi}(b),
where the green strip includes both the NLC QCD SRs uncertainties,
generated by our bunch of the allowed pion DA,
and by the scale-setting ambiguities at the NLO level.

From the phenomenological point of view, 
the most interesting result here is 
that the BMS pion DA~\cite{BMS01} 
(out of a ``bunch'' of similar doubly-peaked endpoint-suppressed pion DAs) 
yields to predictions for the electromagnetic form factor 
very close to those obtained with the asymptotic pion DA.
Conversely, we see that a small deviation 
of the prediction
for the pion form factor 
from that obtained with the asymptotic pion DA
does not necessarily imply that the underlying pion DA 
has to be close
to the asymptotic profile.
Much more important is the behavior of the pion DA in the endpoint
region $x\to 0\,, 1$.

\section{New lattice data and pion DA}
Rather recently new high-precision lattice measurements 
of the pion DA second moment
$\va{\xi^2}_{\pi} = \int_0^1(2x-1)^2\varphi_\pi(x)\,dx$
appeared~\cite{DelD05,Lat06}.
Both groups extracted from their respective simulations, 
values of $a_2$ at the Schmedding--Yakovlev scale
$\mu^2_\text{SY}$ around $0.24$,
but with different error bars.

It is remarkable that these lattice results are in striking agreement
with the estimates of $a_2$ both from NLC QCD SRs~\cite{BMS01} 
and also from the CLEO-data analyses---based 
on LCSR---\cite{SY99,BMS02}, 
as illustrated in Fig.\ \ref{fig:Lat.Ren.CLEO}(a), 
where the lattice results of~\cite{Lat06}
are shown in the form of a vertical strip, 
containing the central value with associated errors.
Noteworthily, the value of $a_2$ of the displayed lattice measurements
(middle line of the strip) is very close to the CLEO best fit 
in~\cite{BMS02} ({\blue\ding{58}}).

\section{Improved model for NLCs and pion DA}
\begin{figure}[h]
 \centerline{\begin{minipage}{0.93\textwidth}\hspace*{-0.013\textwidth}
  \begin{tabular}{cc}
   \begin{minipage}{0.48\textwidth}
    \centerline{\includegraphics[width=\textwidth]{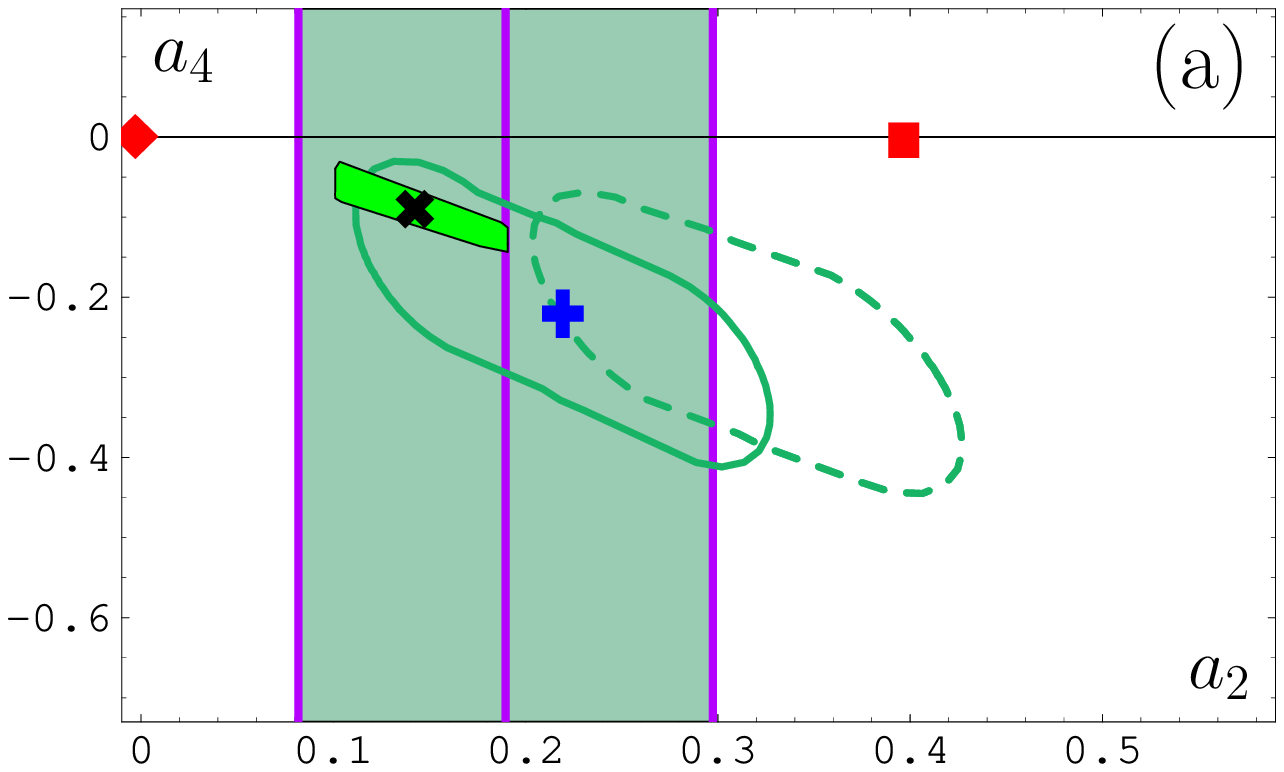}}
   \end{minipage}&
   \begin{minipage}{0.48\textwidth}
    \centerline{\includegraphics[width=\textwidth]{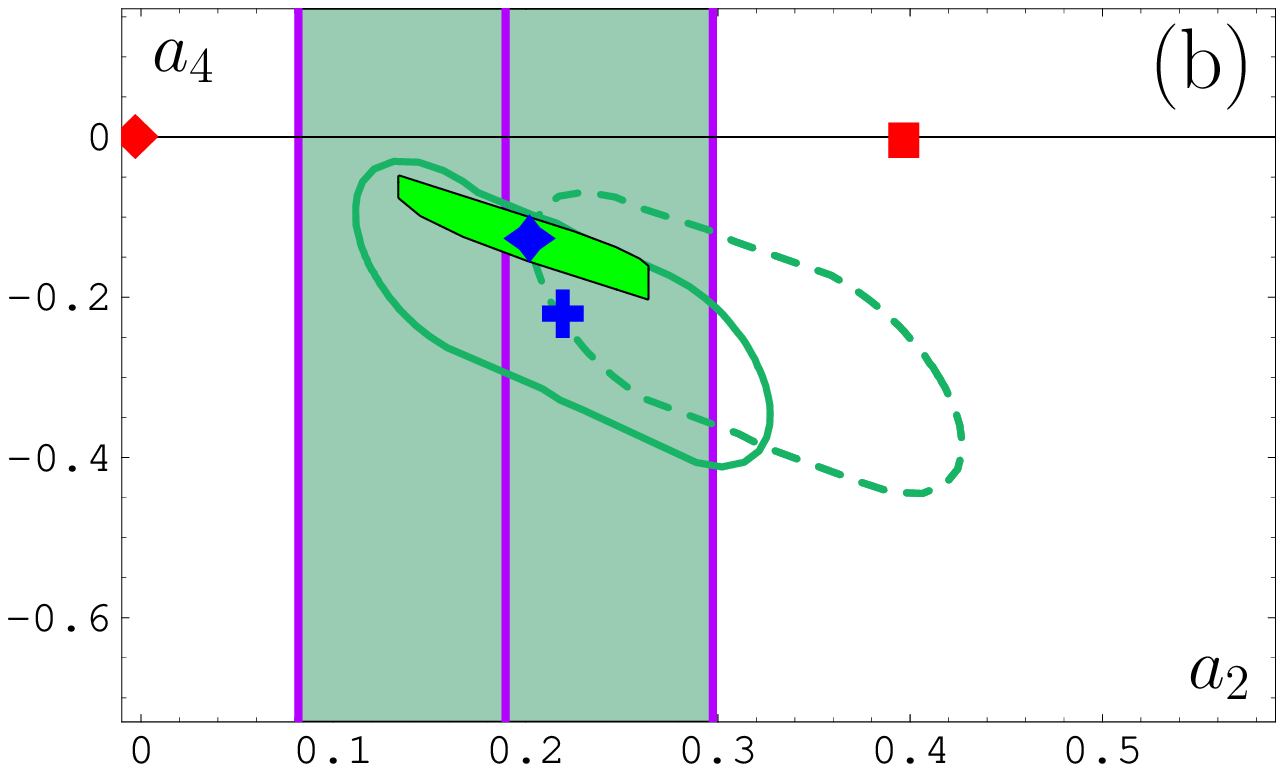}}
   \end{minipage} 
 \end{tabular}
  \caption{The results of the CLEO data analysis for the pion DA
        parameters $a_2$ and $a_4$, evaluated at $\mu^2_\text{SY}=5.76~\text{GeV}^2$.
        The lattice results of \protect\cite{Lat06} are shown for comparison 
        as shaded area, whereas the renormalon-based $1\sigma$-ellipse 
        of~\protect\cite{BMS05lat} is displayed by the green dashed line.
        The shaded strip in panel \textbf{(a)} shows the corresponding results
        for the BMS-``bunch'', whereas in panel \textbf{(b)} --- the new bunch,
        corresponding to the improved Gaussian model of QCD vacuum.
        \label{fig:Lat.Ren.CLEO}}
 \end{minipage}} 
\end{figure}

The quark-gluon-antiquark condensates are usually parameterized
in the following form
\begin{eqnarray}
\va{\bar{q}(0)\gamma_\mu(-g\widehat{A}_\nu(y))q(x)}&=&
       (y_\mu x_\nu-g_{\mu\nu}(yx))\overline{M}_1(x^2,y^2,(y-x)^2)\nonumber\\
      &+&
      (y_\mu y_\nu-g_{\mu\nu}y^2)\overline{M}_2(x^2,y^2,(y-x)^2)\,,\nonumber
\\
\va{\bar{q}(0)\gamma_5\gamma_\mu(-g\widehat{A}_\nu(y))q(x)}&=&
       i\varepsilon_{\mu\nu yx}\overline{M}_3(x^2,y^2,(y-x)^2)\,,
\vspace{-5mm}\nonumber
\end{eqnarray}
with ($A_{1,2,3}=A_0\left(-\frac32,2,\frac32\right)$)
\begin{eqnarray}
 \overline{M}_i(x^2,y^2,z^2)
  = A_i\int\!\!\!\!\int\limits_{\!0}^{\,\infty}\!\!\!\!\int\!\!
        d\alpha \, d\beta \, d\gamma \,
         f_i(\alpha ,\beta ,\gamma )\,
          e^{\left(\alpha x^2+\beta y^2+\gamma z^2\right)/4}\,.
\end{eqnarray}
The minimal model of nonlocal QCD vacuum suggests 
the following Ansatze
\begin{eqnarray}
 \label{eq:Min.Anz.qGq}
  f_i\left(\alpha,\beta,\gamma\right)
   = \delta\left(\alpha -\Lambda\right)\,
       \delta\left(\beta -\Lambda\right)\,
        \delta\left(\gamma -\Lambda\right)
\end{eqnarray}
with $\Lambda=\frac12\lambda_q^2$ and faces problems 
with QCD equations of motion 
and gauge invariance of 2-point correlator of vector currents.
In order to fulfil QCD equations of motion exactly
and minimize non-transversity of $V-V$ correlator
we suggest~\cite{BP06} the improved model of QCD vacuum with 
\begin{eqnarray}
 \label{eq:Imp.Anz.qGq}
 f^\text{imp}_i\left(\alpha,\beta,\gamma\right)
  = \left(1 + X_{i}\partial_{x} + Y_{i}\partial_{y} + Y_{i}\partial_{z}\right)
         \delta\left(\alpha-x\Lambda\right)
          \delta\left(\beta-y\Lambda\right)
           \delta\left(\gamma-z\Lambda\right)\,,
\end{eqnarray}
where $z=y$, $\Lambda=\frac12\lambda_q^2$ and
\begin{subequations}
\begin{eqnarray}
  X_1 &=& +0.082\,;~X_2 = -1.298\,;~X_3 = +1.775\,;~x=0.788\,;~~~\\
  Y_1 &=& -2.243\,;~Y_2 = -0.239\,;~Y_3 = -3.166\,;~y=0.212\,.~~~
\end{eqnarray}
\end{subequations}
Then the NLC sum rules~\ref{eq:NLC.SR.pion.DA} produce~\cite{BP06}
a ``\textit{bunch}'' of 2-parameter pion DA models (\ref{eq:pi.DA.2Geg})
at $\mu^2=1.35$ GeV$^2$.
The coordinates of the central point {\blue\ding{70}} are 
$a_2=0.268$ and $a_4=-0.186$. 
These values correspond to $\langle{x^{-1}}\rangle_\pi^{\text{bunch}} = 3.24\pm0.20$,
which is in agreement with the result of an independent sum rule, viz.,
$\langle{x^{-1}}\rangle_\pi^{\text{SR}}=3.40\pm0.34$.
In order to facilitate possible applications of our ``bunches'' 
we suggest 
the following approximate analytic descriptions of their boundaries ---
we specify upper and lower curves, $a_4^{\pm}(a_2)$, 
and left and right vertical lines, fixed by their abscissa values, $a_2^\text{L,R}$:
\begin{subequations}
\begin{eqnarray}
 \label{eq:BMS.bunch}
  a_4^{\text{BMS},+}(a_2) &=& 0.177 - 2.41 \,a_2 + 7.77\,a_2^2 - 14.0\,a_2^3\,;\\
  a_4^{\text{BMS},-}(a_2) &=&-0.027 - 0.246\,a_2 - 4.12\,a_2^2 + 8.86\,a_2^3\,;\\
  a_2^{\text{BMS, L}}     &=& 0.134\,;\quad 
  a_2^{\text{BMS, R}}    \ =\ 0.251\,,
\end{eqnarray}
\end{subequations}
for the BMS bunch, whereas for the improved bunch 
\begin{subequations}
\begin{eqnarray}
 \label{eq:BP.bunch}
  a_4^{\text{BP},+}(a_2) &=& 0.352 - 4.25\,a_2 + 14.0\,a_2^2 - 18.9\,a_2^3\,;\\
  a_4^{\text{BP},-}(a_2) &=& 0.588 - 6.89\,a_2 + 20.7\,a_2^2 - 23.6\,a_2^3\,;\\
  a_2^{\text{BP, L}}     &=& 0.177\,;\quad 
  a_2^{\text{BP, R}}    \ =\ 0.346\,.
\end{eqnarray}
\end{subequations}
Allowed values of both bunch parameters $a_2$ and $a_4$
after NLO-evolution to $\mu^2=5.76$ GeV$^2$ 
are shown in Fig.\ \ref{fig:Lat.Ren.CLEO} in a form of 
shaded slanted rectangles around central points \ding{54}
and {\blue\ding{70}}.

We emphasize here that BMS model~\cite{BMS01},
shown in Fig.\ \ref{fig:Lat.Ren.CLEO}(a) by symbol {\ding{54}},
is inside the allowed region dictated by the improved QCD vacuum model.
This means that all the characteristic features of the BMS bunch 
are also valid for the improved bunch:
NLC-dictated models are end-point suppressed,
although are double-humped.

We see in Fig.\ \ref{fig:Lat.Ren.CLEO}(b) 
that the improved bunch~\cite{BP06} is 
even in a better agreement with the recent lattice results~\cite{Lat06},
shown in the form of a vertical strip, 
containing the central value with associated errors.
Remarkably, the value of $a_2$ of the displayed lattice measurements
(middle line of the strip) is very close to the central point 
of the ``improved bunch'' ({\blue\ding{70}}), 
whereas the whole bunch,
dictated by the improved NLC QCD SRs~\cite{BP06},
is inside the strip\footnote{%
The same is valid for the BMS ``bunch'' as well, as can be seen in Fig.\ \ref{fig:Lat.Ren.CLEO}(a).}
and completely inside the standard CLEO $1\sigma$-ellipse.

\section{Conclusions}
Let me conclude with the following
observations:
\begin{itemize}
\item  NLC QCD SR method for the pion DA gives us 
the admissible bunches of DAs
for each value of $\lambda_q$.

\item NLO LCSR method produces new constraints
on the pion DA parameters ($a_2$ and $a_4$)
in conjunction with the CLEO data.

\item Comparing results of the NLC SRs
with new CLEO constraints allows to fix
the value of QCD vacuum nonlocality: $\lambda_q^2=0.4~\gev{2}$.

\item This bunch of pion DAs agrees well 
E791 data  on diffractive dijet $\pi+A$-production, 
with JLab F(pi) data on the pion form factor, 
and with recent lattice data. 

\item Taking into account QCD equations of motion
 for NLCs and transversity of vacuum polarization
 allows us to put the pion DA bunch just inside $1\sigma$-ellipse
 of CLEO-data constraints.  

\end{itemize}

\section*{Acknowledgments}
This investigation was supported in part 
by the Heisenberg--Landau Programme, grant 2006, 
and the Russian Foundation for Fundamental Research, 
grants No.\ 06-02-16215.
   


\end{document}